\documentclass[10pt,a4paper,twocolumn]{article}
\usepackage{fukugo}
\usepackage{amsmath}
\usepackage{amsfonts}
\usepackage{amssymb}
\newcommand{\beq}{\begin{equation}}
\newcommand{\eeq}{\end{equation}}
\newcommand{\bea}{\begin{eqnarray}}
\newcommand{\eea}{\end{eqnarray}}

\begin{document}
%
\title{An Attempt to Study Pentaquark Baryons in String Theory}
%
%
%

\author{{\large{Akio SUGAMOTO}} \medskip \\
   \it{ Department of Physics and Chemistry,} \\
   \it{ Graduate School of Humanities and Sciences,       } \\
   \it{ Ochanomizu University                             }
        }
%
%
%
\maketitle
\pagestyle{empty}
\thispagestyle{empty}
\setlength{\baselineskip}{13pt}
%
%

%
Abstract
\hrulefill

An attempt to study recently observed Pentaquark baryons is performed in the 
dual string theory of QCD.  Mass formulae for Pentaquark baryons are naively estimated in the Maldacena's prototype model for supersymmetric QCD and a more realistic model for the ordinary QCD.  

\hrulefill

In this talk we attempt to study Pentaquark baryons in the dual string theory of QCD, based on a work done in collaboration with M. 
Bando, T. Kugo, and S. Terunuma [BKST].  (The talk is given at 2nd international symposium on New Developments of Integrated Sciences held at Ochanomizu University on March 16 (2004).)

The first Pentaquark baryon 
$\Theta^{+}$ was found at SPring-8 [SP8] using 
$\gamma+n\rightarrow K^{-}\Theta^{+}\rightarrow K^{-}K^{+}n$.  $\Theta^{+}$ 
is considered to be $(ud)(ud)\bar{s}$.  The mass is $M(\Theta^{+})=1540 \pm 
10 \mbox{MeV}$, and the width  $\Gamma(\Theta^{+}) \le 25 \mbox{MeV}$.  The other 
penta-quark $\Xi^{--}((ds)(ds)\bar{u})$ was subsequently observed at  CERN 
[NA49] with mass $M(\Xi^{--})=1862 \mbox{MeV}$, and width $\Gamma(\Xi^{--}) \le 18 
\mbox{MeV}$.  Prior to these experiments, Pentaquarks were predicted as a chiral soliton by 
Diakonov {\it et al.} [DPP] with reasonable values for mass and 
width of $\Theta^{+}$.  The Diquark model by Jaffe and Wilczek [JW] is helpful, since 
if a Diquark (ud) is considered as an anti-scalar quark 
($\bar{\tilde{s}}$),  $\Theta^{+}$ becomes 
$\bar{\tilde{s}}_1\bar{\tilde{s}}_2 \bar{s}_3$ which is similar to anti-$\Omega^{-}$.    

When we draw a picutre of $\Theta^{+}$ 
as quarks conected by colored strings, we find a beautiful shape. 
The picture is as follows:
The colored strings (colored fluxes) springing out from u- and d-quarks terminate at a junction.  We have two Diquark pairs, $(u, d)_1$ and $(u, d)_2$, so we have two junctions, $J_1$ and $J_2$. The other junction $J_0$ is to be prepared, from which three colored strings come out and terminate to $J_1$, $J_2$ and the anti-s-quark.  The shape looks like a branched web.
In this colored string picture, the mass of $\Theta^{+}$ is to be estimated as the total length of 
strings.  We have a hope that Pentaquark having this shape is difficult to decay into meson and nucleon, since the recombination of strings associated with their splitting, or the pair production of quarks is required in the intermediate states.  

To produce colored strings, however, 
non-perturbative QCD is compulsory.  Owing to J. Maldacena [M], we can now replace the non-perturbative QCD by a classical 
gravity theory, being dual to QCD.  As usual we prepare $N_{c}$ 
coincident colored "D3-branes", or "3 dimesionally extended membranes", and open strings connecting two of these colored branes are gluons. To 
introduce quarks with flavors, we add $N_{f}$ flavored D7 branes [KK].  Now quark is an open stirng 
connecting a colored brane and a flavored brane.  The intrinsic 
mass of quark is the energy stored inside the minimum length of open 
string, so the flavored brane may be separeted from the colored 
branes by a distance $U$, and the intrinsic quark mass is $U\times 
\mbox{(string tension)} =U/2\pi\alpha'$.

First we study Pentaquarks in the prototype model by Maldacena [M].  This is the dual gravity model 
given in the curved space of $AdS_5\times S^5$, a natural space bent by 
the mass and the charge of $N_{c}$ colored branes.  In the 
presence of flavored branes, the model becomes N=2 
supersymmetric gluodynamics with $N_{f}$ hypermultiplets of 
quarks which is, however,  not a realistic model.  To study more realistic models, we have to refer to the recent works with the breaking of supersymmetry [N=0a][N=0b][N=0d][N=0c].

In the former prototype model, the background metric of the four dimensional space-time plus one extra dimension $U$ reads
\begin{equation}
ds^2=f(U)(-dt^2+dz^2+d{\bf x}_{\perp}^2)+g(U)dU^2,
\end{equation}
where $f(U)=g(U)^{-1}=(U/R)^2$, and the radius of $AdS_5$ is given by the coupling $g_{s}$ of closed string, or QCD coupling $\alpha_{c}$ as 
$R^4=4\pi g_{s}N_{c}=8\pi \alpha_{c} N_{c}$. In this paper every variable is made dimensionless by multiplying proper powers of $\alpha'$ of $(\mbox{length})^2$.

The string action under the curved space of Eq.(1) is given by
\begin{equation}
S= \frac{1}{2\pi} \int d\tau d\sigma 
\sqrt{-(\dot{X}^{M}\dot{X}_{M})(X'^{N}X'_{N})+(\dot{X}^{M}X'_{M})^2},
\end{equation}
where $X^{M}(\tau,\sigma)$ describes the world sheet of string, and $\dot{}$ and $'$ are derivatives by $\tau$ and $\sigma$, respectively.  Choose $\sigma=z$ and $\tau=t$, then in the static limit the action becomes $S=\frac{\Delta t}{2\pi}\int dz L$, where
\begin{equation}
L=\sqrt{f(U)^2(1+({\bf x}'_{\perp})^2))+f(U)g(U)(U')^2}.
\end{equation}
If we consider $z$ as "time", we have three conserved quantities, "energy" $H$, and perpendicular "momenta" ${\bf p}_{\perp}$.  Then the perpendicular coordinates move uniformely in "time", that is, ${\bf x}_{\perp}/z={\bf p}_{\perp}/(-H)$.  Now, the usual energy stored inside string is estimated by
\begin{equation}
E=\frac{1}{2\pi} \int^{U_2}_{U_1}du \sqrt{ \frac{f(u)^3g(u)}{f(u)^2-(-H)^2-({\bf p}_{\perp})^2}},
\end{equation}
\begin{equation}
\frac{z}{(-H)}=\int^{U_2}_{U_1}du \sqrt{ \frac{g(u)}{f(u)(f(u)^2-(-H)^2-({\bf p}_{\perp})^2)}}.
\end{equation}

In the picture of Pentaquark we have three junctions of strings, $J_0, J_1$, and $J_2$.  At $J_0$ three strings are assumed to separate with opening angles $2\pi/3$ and $U'=0$.  At $J_1$(or $J_2$), string (1) from $J_0$, having coordinates $({\bf x}_{\perp}=({\bf 0}, z, U^{(1)}(z))$, is assumed to split into string (2) with $({\bf x}, z, U^{(2)}(z))$ and string (3) with $(-{\bf x}, z, U^{(3)}(z))$.  In the locally flat space near $J_1$ (or $J_2$), three strings lie on a single plane, pulling each other with equal strength, that leads to the opening angles between strings be all $2\pi/3$.  Then, we obtain the following connection conditions at the junction,
\begin{equation}
(-H)_{(2)}=(-H)_{(3)}=\frac{1}{2} (-H)_{(1)},
\end{equation}
\begin{equation}
({\bf x}'_{\perp})^2_{(2), (3)}=3\left(1+\frac{g(U)}{f(U)}(U'_{(1)})^2\right).
\end{equation}

With these conditons, it is straightforward to evaluate the static energy of Pentaquarks in terms of coordinates of quarks.  Here we adopt a simple approximation in which $U$ is largely separated, namely, $U(J_0)\ll U(J_{1,2})\ll U((u,d)_{1,2})$ and $U(J_0)\ll U(\bar{s}_3)$.  Then, we obtain
\begin{equation}
E(\Theta^{+})=m_s+2(m_u+m_d)+V(z_{1,2,3}; x_{1,2}), 
\end{equation}
\begin{equation}
V=-\frac{aR^2}{2\pi}\left(\frac{1}{z_1}+\frac{1}{z_2}+\frac{1}{z_3}\right)-\frac{bR^2}{2\pi}\left(\frac{1}{x_1}+\frac{1}{x_2}\right),
\end{equation}
where $z_{1,2,3}$ are the center of mass coordinates (distances from $J_0$) of $(ud)_{1,2}$ and $\bar{s}_3$, while $x_{1,2}$ are relative coordinates inside Diquarks, $(ud)_{1,2}$.  The calculable constants, $a=0.359$ and $b=0.236$ depend on the adopted dual string model.

Solving the non-relativistic Sch\"odinger equation, we obtain the mass of $\Theta^{+}$ as the lowest eigenvalue,
\begin{equation}
M(\Theta^{+})=2(m_u+m_d)(1-A-B)+m_s(1-A),
\end{equation}
where $(A, B)=\alpha_c N_c (a^2, b^2)/\pi$.
Similarly, we have for $\Xi^{--}$
\begin{equation}
M(\Xi^{--})=2(m_d+m_s)(1-A-B)+m_u(1-A).
\end{equation}
The mass of Triquark family of nucleon is simply $M(N)=(m_1+m_2+m_3)(1-A)$.

Although our prototype model and approximation are very naive, we apply the constituent quark masses of $m_u=m_d=360 \mbox{MeV}$ and $m_s=540 \mbox{MeV}$ for the quark masses and take $N_{c}=3$, we obtain the values of $\alpha_c$ so as to reproduce the observed masses of Pentaquarks and Triquarks.  For $\Theta^{+}$, $\Xi^{--}$, we have 1.38, 0.79 for $\alpha_c$, while for Triquarks, $N$, $\Sigma$, and $\Xi$, we have 1.07, 0.43, and 0.69, respectively.  These values are not too bad, in comparison with the observed value $\alpha_c=0.35 \pm $0.03 at $M_{\tau}=1777 \mbox{MeV}$.  
Therefore, this direction of study on Pentaquarks are worth persuing.

If we go to a more realistic model with $N_{c}$ colored D4 branes and $N_{f}$ flavored D6 branes [N=0a][N=0b][N=0d][N=0c], we may have the following potential including linear potential:
\begin{equation}
V=k(z_1+z_2+z_3)+\sum_{i=1,2}\left(V(x_{i})+V(x'_{i})\right),
\end{equation}
where $k=(\frac{2}{3})^{3} \alpha_{c} N_{c} (M_{KK})^2$, \begin{equation}
V(x) \sim -\frac{a}{M} \frac{1}{x^2}, 
\end{equation}
for $x \rightarrow \infty$, and
\begin{equation}
V(x) \sim  -\frac{1}{2}m+\frac{2}{\sqrt{3}}(m^3 M)^{\frac{1}{2}}x, 
\end{equation}
for $x \rightarrow 0$,  where $a=0.0689$.  $M=M_{KK} /(\alpha_{c} N_{c})$ is defined with the mass scale $M_{KK}$ of Kaluza-Klein compactification which is necessary to break supersymmetry and to obtain the more realistic model for QCD [N=0a][N=0b][N=0d][N=0c].
 
Then, we obtain the following mass formulae in a naive approximation:
\begin{eqnarray}
&&\bar{M}(\mbox{Pentaquark}) \nonumber \\
&=& \bar{m}_1+\bar{m}_2+\bar{m}_3  \nonumber\\
&+&\frac{2}{3}(\alpha_{c} N_{c})^2 \{ 2 (\bar{m}_1+\bar{m}_2)^{-\frac{1}{3}} +(\bar{m}_3)^{-\frac{1}{3}} \} \nonumber \\
&+&  6^{\frac{2}{3}}\{(\bar{m}_1)^{\frac{2}{3}}+(\bar{m}_2)^{\frac{2}{3}} \},
\end{eqnarray}
and 
\begin{eqnarray}
& &\bar{M}(\mbox{Triquark}) \nonumber \\
&=&\bar{m}_1+\bar{m}_2+\bar{m}_3  \nonumber\\
&+&\frac{2}{3}(\alpha_{c} N_{c})^2 \sum_{i=1-3} (\bar{m}_{i})^{-\frac{1}{3}} .
\end{eqnarray}

Here, Pentaquark is considered to consist of five quarks $((q_1 q_2)^2 \bar{q}_3)$, while Triquark is $(q_1 q_2 q_3)$.  All the masses with bar are those normalized by $M$.  We have to compare these formulae with the experimental date so far obtained.  

In conclusion, we want to say that the discovery of Pentaquark baryons may give us a good opportunity to start the "stringy integrated sciences" combining  the nuclear physics and the high energy physics, since the string theory is now coming down to the real world.  

[Noe added] After the symposium, numerical calculation was completed in the QCD like model, giving the following results;
$m_u=m_d=$313-312 MeV, $m_s=$567-566 MeV, and  $M(\Theta)=$1,577-1,715 MeV, $M(\Xi)=$1,670-1,841 MeV, for $\alpha_c$=0.33 and KK mass scale $M=M_{KK}=$2-5 MeV.

The details will be described in [BKST].

%

\end{document}